# Incommensurate phases in statistical theory of the crystalline state

V. A. Golovko

Moscow State Evening Metallurgical Institute

Lefortovsky Val 26, Moscow 111250, Russia

E-mail: mgvmi-mail@mtu-net.ru

Abstract

The paper continues a series of papers devoted to treatment of the crystalline state on the basis of the approach in equilibrium statistical mechanics proposed earlier by the author. This paper is concerned with elaboration of a mathematical apparatus in the approach for studying second-order phase transitions, both commensurate and incommensurate, and properties of emerging phases. It is shown that the preliminary symmetry analysis for a concrete crystal can be performed analogously with the one in the Landau phenomenological theory of phase transitions. The analysis enables one to deduce a set of transcendental equations that describe the emerging phases and corresponding phase transitions. The treatment of an incommensurate phase is substantially complicated because the symmetry of the phase cannot be described in terms of customary space groups. For this reason, a strategy of representing the incommensurate phase as the limit of a sequence of long-period commensurate phases whose period tends to infinity is worked out. The strategy allows one to obviate difficulties due to the devil's staircase that occurs in this situation.

# 1. Introduction

Incommensurate phases in dielectric crystals were intensively investigated both theoretically and experimentally in the 1970–80s [1]. Interest in them persists nowadays as well. The incommensurate phases are a special state of solids that differs from ordinary crystals by the lack of periodicity in one or more directions whereas a long-range order exists unlike amorphous solids. Usually, the incommensurate phases attend some second-order phase transitions: an incommensurate phase can appear as an intermediate phase between a high-temperature parent phase and a low-temperature (commensurate) phase. The reasons for the appearance of the incommensurate phase and its properties are well described by the Landau theory of phase transitions. The Landau theory, however, is a phenomenological theory relying upon some general argumentation, the theory that is not designed to explain microscopic causes of the phenomenon, i.e., causes relevant to properties of molecules or atoms.

A strict microscopic theory is statistical mechanics inasmuch as it tries and explains phenomena by proceeding from a knowledge of the intermolecular potential and of other properties of the molecules and atoms. In Ref. [2], a new approach in equilibrium statistical mechanics was worked out which is not based upon the Gibbs ensembles (see also [3]). The classical version of the approach used in the present paper (the approach was devised initially for quantum systems) leans upon the BBGKY hierarchy of equations for reduced distribution functions while the distinctive feature of the approach lies in constructing thermodynamics compatible with the hierarchy. Among other things, the approach proved to be seminal in studies of the crystalline state [4-6]. In particular, in Ref. [5] the approach was employed for studies of second-order phase transitions, and the results obtained are in full accord with the Landau phase transition theory.

The main aim in this paper is to apply the statistical approach to investigation of the incommensurate phases in dielectric crystals. Besides, the example of a second-order phase transition considered in [5] is rather peculiar, in which the possibility of the transition is directly seen from the expressions for effective potentials deduced in that paper. Usually, an individual analysis is required to see the possibility of a phase transition in a concrete crystal. We shall show also how the analysis can be carried out in statistical theory. This can be done analogously with the Landau theory although some refinements are needed which are not obligatory in the Landau theory.

Many theoretical studies on incommensurate phases are based on the Landau free energy characteristic of ammonium fluoberyllate $(NH_4)_2BeF_4$ whose high-temperature space group is $D_{2h}^{16}$ (note that the high-temperature phase of many crystals that have an incommensurate phase pertains to the same orthorhombic space group). To be specific and for the purposes of comparison with the Landau theory we shall carry our investigation using, as an example, the sequence of the phase transitions that occur in ammonium fluoberyllate. At the same time, the approach that will be employed is developed in detail only for systems containing particles of one kind with spherically symmetric interaction.[1] An equilibrium crystal composed of such particles should be of cubic symmetry. Of course, the orthorhombic lattice may be obtained from the cubic one by applying appropriate external stresses. However, space group $D_{2h}^{16}$ is not a subgroup of any other space group of a higher crystal class if no change in the number of particles in the unit cell is involved [8] and thereby it cannot be obtained from a cubic space group by continuous deformation. For this reason, a $D_{2h}^{16}$ crystal composed of particles of one kind with spherically symmetric interaction can be in a metastable state alone, which may also entail other peculiarities irrelevant to our problem. Our main goal in this paper is to work out a mathematical apparatus for treating incommensurate phases in statistical theory, and the example of space group $D_{2h}^{16}$ is sufficiently general and well suited for this. At the same time, it should be emphasized that the results obtained cannot be applied directly to ammonium fluoberyllate whose molecules are by no means spherically symmetric.

As mentioned above, one of the aims in the present paper is to compare the approach used and the results obtained with its help with the ones that follow from the Landau phase transition theory. To this end, in Section 2 we outline the Landau theory as applied to incommensurate phases and adduce its results needed in this paper. In Section 3, we formulate the basic equations of the statistical approach employed for consideration of the crystalline state and apply them in Section 4 to the $D_{2h}^{16}$ parent phase from which the phase transitions studied commence. We show in Section 5 how the symmetry analysis concerning a commensurate phase transition can be carried out in statistical theory and how the theory describes the emerging phase. Section 6 is devoted to the incommensurate phases in statistical theory and to corresponding phase transitions.

**2. Incommensurate phases in the Landau phase transition theory**

A typical density of the Landau free energy used for treatment of an incommensurate phase is of the form [9]

---

[1] Extension of the approach to systems that contain particles of several kinds can be found in [7].

$$\tilde{F}(y) = \overline{F} + \frac{\alpha}{2}\left(\eta^2 + \xi^2\right) + \frac{\beta_1}{4}\left(\eta^2 + \xi^2\right)^2 + \frac{\beta_2}{4}\left[\left(\eta^2 - \xi^2\right)^2 - (2\eta\xi)^2\right] + \sigma\left(\xi\frac{d\eta}{dy} - \eta\frac{d\xi}{dy}\right)$$

$$+ \frac{\delta}{2}\left[\left(\frac{d\eta}{dy}\right)^2 + \left(\frac{d\xi}{dy}\right)^2\right]. \tag{2.1}$$

Here $\overline{F}$ is a term irrelevant to our problem, $\eta$ and $\xi$ are components of the order parameter. The term with the coefficient $\sigma$ is the Lifshitz invariant (for convenience in the following we assume that the incommensurate modulation occurs along the *y*-axis). The invariant can exist only if the order parameter has two components or more. In case the invariant is admitted by the irreducible representation that describes the phase transition in question, there inevitably appears an incommensurate phase. The invariants of fourth order in $\eta$ and $\xi$ with the coefficients $\beta_1$ and $\beta_2$ are written in a form convenient for further treatment (the simplest form of the quartic invariants in this case is $\eta^4 + \xi^4$ and $\eta^2\xi^2$). We assume that $\beta_1 > |\beta_2|$, otherwise invariants of sixth order in $\eta$ and $\xi$ must be taken into account. If the length of the crystal in the *y*-direction is $L$, the free energy per unit length is

$$F = \frac{1}{L}\int_0^L \tilde{F}(y)dy. \tag{2.2}$$

As is usual in the Landau theory, we presume that the coefficient $\alpha$ alone depends upon the temperature varying linearly with it.

In Section 5 we shall demonstrate how the free-energy density of (2.1) can be obtained in the case of ammonium fluoberyllate. It should be observed that the Landau free energy may be readily written down once one knows the number of different invariants because their form can be established with ease from simple considerations.

Were the Lifshitz invariant lacking, the minimum of $F$ would correspond to a uniform solution $\eta$ = constant and $\xi$ = constant because of the last term in (2.1) which is always present (we assume that $\delta > 0$). Minimizing the free energy in that event, one would, if $\alpha < 0$, arrive at two nontrivial solutions corresponding to two possible commensurate phases:

$$(1)\ \eta \neq 0,\ \xi = 0 \text{ or vice versa},\quad \eta^2 = -\frac{\alpha}{\beta_1 + \beta_2},\quad F = \overline{F} - \frac{\alpha^2}{4(\beta_1 + \beta_2)}; \tag{2.3}$$

$$(2)\ \eta^2 = \xi^2 = -\frac{\alpha}{2(\beta_1 - \beta_2)},\quad F = \overline{F} - \frac{\alpha^2}{4(\beta_1 - \beta_2)}. \tag{2.4}$$

The first solution is energetically preferable if $\beta_2 < 0$, the second one if $\beta_2 > 0$.

Before proceeding further it is instructive to discuss the problem from the viewpoint of the concept of a soft mode in the high-temperature parent phase. The soft mode in ammonium fluoberyllate occurs at the boundary of the Brillouin zone to which the wavevector $K = a_2/2$ corresponds, where $\mathbf{a}_2$ is the basic reciprocal-lattice vector oriented along the y-axis in the present notation (Fig. 1 where the Brillouin zone boundary is denoted as point B with the vertical line). If the Lifshitz invariant does not exist at $K = a_2/2$, the soft-mode branch has an extremum at $K = a_2/2$ as in Fig. 1(a). In case the extremum is a minimum (curve 1 in Fig. 1(a)), as the temperature lowers (the length of the segment AB in Fig. 1 is proportional to the coefficient $\alpha$ in (2.1)), it is the frequency $\omega$ at $K = a_2/2$ that vanishes. As a result, in the crystal an atom-displacement wave is frozen-in and there occurs a phase transition at $\alpha = 0$ into a commensurate phase described by (2.3) or (2.4) whereas the crystal period along the y-axis doubles for the wavelength corresponding to $K = a_2/2$ is equal to $2d_2$ where $d_2$ is the period in the parent phase.

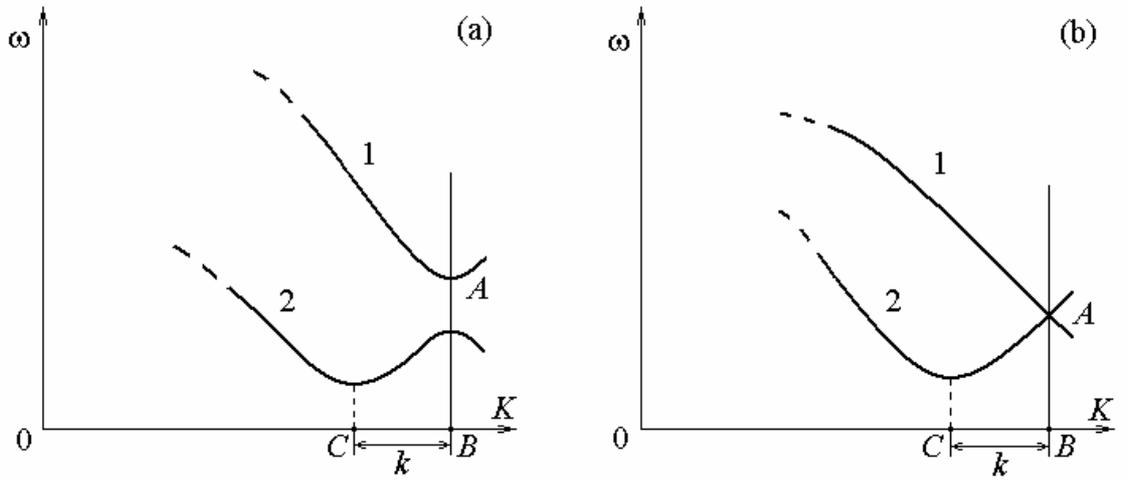

**Fig. 1**. Dispersion curves (phonon branches) near the Brillouin zone boundary (point B) in the parent phase: (a) the Lifshitz invariant is lacking, (b) the Lifshitz invariant exists.

The existence of the Lifshitz invariant at $K = a_2/2$ signifies a non-zero slope of the phonon branch at point A (Fig. 1(b)). In actual fact, two phonon branches meet at point A in this instance as long as all dispersion curves can be transferred to the first Brillouin zone (by the way, this explains why the order parameter in this case has two components or more). The lower branch is necessarily a minimum inside the zone and, as the temperature lowers, it is the frequency at point C that vanishes first although $\alpha > 0$. As a result, a superstructure with the wavevector $K = a_2/2 - k$ arises in the crystal whereas there is no reason for the ratio $k/a_2$ to be rational. The ratio being irrational, one will have two mutually incommensurable periods in the crystal, namely, the period $d_2$ of the underlying structure that remains and the period of the superstructure, that is to say, one will have an incommensurate phase without any periodicity along the y-axis but with a definite long-range order.

Even if the Lifshitz invariant does not exist, the extremum at $K = a_2/2$ can be a maximum (curve 2 in Fig. 1(a)). As the temperature lowers, it is the frequency at point $C$ that vanishes first, and one will again have an incommensurate phase called type II (the incommensurate phase discussed previously is called type I). The emergence of the incommensurate phase of type II is due to peculiarities of intermolecular interactions that lead to a dispersion curve with a minimum inside the Brillouin zone while the inevitability of the type-I incommensurate phase can be forecast from symmetry considerations. Both of these types are observed experimentally in crystals [1].

We revert now to the Landau free energy. Substituting (2.1) into (2.2) and minimizing the functional $F$ yields the following Euler-Lagrange equations

$$\delta \frac{d^2\eta}{dy^2} + 2\sigma \frac{d\xi}{dy} - \alpha\eta - \beta_1\eta(\eta^2 + \xi^2) - \beta_2\eta(\eta^2 - 3\xi^2) = 0, \tag{2.5}$$

$$\delta \frac{d^2\xi}{dy^2} - 2\sigma \frac{d\eta}{dy} - \alpha\xi - \beta_1\xi(\eta^2 + \xi^2) - \beta_2\xi(\xi^2 - 3\eta^2) = 0. \tag{2.6}$$

Near the phase transition point where $\eta$ and $\xi$ are small, one can neglect the nonlinear terms in (2.5) and (2.6). The resulting linear equations admit a solution

$$\eta = \rho \cos ky, \quad \xi = \rho \sin ky \tag{2.7}$$

with $\rho$ = constant and

$$k = \frac{1}{\delta}\left(\sigma \pm \sqrt{\sigma^2 - \alpha\delta}\right). \tag{2.8}$$

The last equation shows that the maximal value of $\alpha$ at which a real value of $k$ exists is $\alpha = \alpha_0 \equiv \sigma^2/\delta$ while then $k = \sigma/\delta$. Therefore, at $\alpha = \alpha_0 > 0$ there occurs a phase transition into an incommensurate phase with $k = \sigma/\delta$ in accord with Fig. 1(b). If one places (2.7) in (2.1) and (2.2), a term with $\cos 4ky$ disappears after integrating over $y$ and implying that $L \to \infty$. Minimizing the resulting $F$ with respect to $\rho$ gives that near the phase transition point

$$\rho^2 = \frac{\alpha_0 - \alpha}{\beta_1}, \quad F = \overline{F} - \frac{(\alpha_0 - \alpha)^2}{4\beta_1}. \tag{2.9}$$

Eqs. (2.5) and (2.6) do not lend themselves to analytical solution in the general case. The equations were solved numerically [10,11], there exists a strict analytical solution valid for an especial relation between the coefficients in (2.1) [12], approximate methods were exploited as well [13]. The solutions obtained show that, as the temperature lowers the value of $k$ decreases and tends to zero. This tendency of $k$ is understandable from the physical point of view because the free energy contains terms that favour the periodic structure. When $k$ approaches zero, there occurs a phase transition to one of the commensurate phases described by Eqs. (2.3) or (2.4). In the case of ammonium fluoberyllate, the space group of the emerging phase is $C_{2v}^9$. This phase transition referred to as a lock-in transition has specific features in comparison with an ordinary second-order transition [1, Vol.1, Chap. 2].

Near the lock-in transition, the incommensurate phase becomes domain-like. The structure of the incommensurate phase inside the domains factually coincides with that of the corresponding low-temperature commensurate phase whereas distinctions take place only in the domain walls called the discommensurations. If the lock-in transition is continuous, it happens when the distance between the discommensurations becomes infinite, in which case the crystal will contain only one or two domains with the commensurate phase.

Closing the section it should be emphasized that the Landau phase transition theory is strict only in the immediate vicinity of the phase transition point (excluding the critical region [14]). All modes in the crystal are coupled with one another, which gives rise to secondary order parameters that should be taken into account in the free energy and whose role augments as the temperature lowers. In the case of an incommensurate phase, higher derivatives of the order parameters and their powers can play a part if the temperature is not sufficiently close to the phase transition temperature. Appreciably below this temperature, the Landau theory based upon a simplified free energy of the type (2.1) should be regarded only as a model.

## 3. Basic equations of statistical theory of the crystalline state

In the present paper, we assume the same form of the pair correlation function as in [6] and utilize the same equations for the crystalline state. The first equation of the BBGKY hierarchy with that pair correlation function yields, for the singlet density distribution $\rho(\mathbf{r})$,

$$\rho(\mathbf{r}) = C \exp[-U(\mathbf{r})/\theta], \qquad (3.1)$$

where $\theta$ is the temperature in units of energy. The constant $C$ is to be found from the normalization condition

$$\int_V \rho(\mathbf{r})\,d\mathbf{r} = N, \qquad (3.2)$$

where the integration is carried out over the volume $V$ of the crystal that contains $N$ particles.

Eq. (3.1) relates the density $\rho(\mathbf{r})$ with the effective potential

$$U(\mathbf{r}) = \int K_g(|\mathbf{r}-\mathbf{r}'|)\rho(\mathbf{r}')d\mathbf{r}', \qquad (3.3)$$

in which

$$K_g(r) = \int_\infty^r \frac{dK(r')}{dr'} g(r')\,dr'. \qquad (3.4)$$

Here $g(r) \equiv g(|\mathbf{r}|)$ is the pair correlation function and we imply that the particles interact by means of a two-body potential $K(|\mathbf{r}_i - \mathbf{r}_j|)$. It should be remarked that in the present paper, as in Ref. [6], we assume the pair correlation function to be spherically symmetric. We shall return to this question in the concluding section.

Upon placing (3.3) in (3.1) we obtain a nonlinear integral equation for $\rho(\mathbf{r})$. Of course, we should know the pair correlation function $g(r)$ that figures in (3.4). Strictly speaking, the function has to be found from the second equation of the BBGKY hierarchy. In this paper we shall carry out our investigation without specifying the form of $g(r)$. Certainly, to apply results obtained in the paper to a concrete crystal a knowledge of $g(r)$ is required, which will be discussed in the concluding section. Refs. [4–6] demonstrate that many interesting results for a crystal can be obtained without specifying the form of $g(r)$. This is connected with the fact that the leading role for the crystal is played by the first BBGKY equation that is satisfied identically for a fluid where, for this reason, the leading role goes over to the second BBGKY equation for the pair correlation function whereas this second equation is auxiliary for the crystal [5, 6]. In any case, $K_g(r)$ of (3.4) can be regarded as an effective intermolecular potential.

The main idea of Refs. [4–6] as to treating the integral equation for the density $\rho(\mathbf{r})$ is to expand $\rho(\mathbf{r})$ in a Fourier series:

$$\rho(\mathbf{r}) = \sum_{l,m,n=-\infty}^{\infty} a_{lmn} e^{i\mathbf{A}\mathbf{r}}, \qquad (3.5)$$

where $\mathbf{A} = l\mathbf{a}_1 + m\mathbf{a}_2 + n\mathbf{a}_3$ with the basic reciprocal-lattice vectors $\mathbf{a}_1$, $\mathbf{a}_2$ and $\mathbf{a}_3$. Upon substituting (3.5) into (3.3) one finds the expansion for $U(\mathbf{r})$

$$U(\mathbf{r}) = \sum_{l,m,n} a_{lmn} \sigma(A) e^{i\mathbf{A}\mathbf{r}}, \qquad (3.6)$$

in which $A = |\mathbf{A}|$ and

$$\sigma(A) = \int K_g(|\mathbf{r}|) e^{i\mathbf{A}\mathbf{r}} d\mathbf{r} = \frac{4\pi}{A} \int_0^{\infty} r K_g(r) \sin Ar \, dr. \qquad (3.7)$$

If one inserts (3.5) and (3.6) into (3.1), one obtains a set of equations for $a_{lmn}$ [4]:

$$a_{lmn} = \frac{\rho_0}{8\pi^3 G} \int_0^{2\pi}\int_0^{2\pi}\int_0^{2\pi} \exp\left[-\frac{1}{\theta}\sum_{l',m',n'=-\infty}^{\infty}{}' a_{l'm'n'} \sigma(A') e^{i(l'\xi_1 + m'\xi_2 + n'\xi_3)} - i(l\xi_1 + m\xi_2 + n\xi_3)\right]$$

$$\times d\xi_1 d\xi_2 d\xi_3 \qquad (3.8)$$

with

$$G = \frac{1}{8\pi^3} \int_0^{2\pi} \int_0^{2\pi} \int_0^{2\pi} \exp\left[-\frac{1}{\theta} \sum_{l,m,n}{}' a_{lmn}\, \sigma(A)\, e^{i(l\xi_1 + m\xi_2 + n\xi_3)}\right] d\xi_1\, d\xi_2\, d\xi_3, \qquad (3.9)$$

where the prime over the summation signs denotes omission of the term with $l' = m' = n' = 0$ in (3.8) or with $l = m = n = 0$ in (3.9), $A' = |\mathbf{A}'| = |l'\mathbf{a}_1 + m'\mathbf{a}_2 + n'\mathbf{a}_3|$, and $\rho_0 = a_{000} = N/V$ is the average number density. It will be noted that Eqs. (3.8) and (3.9) are written down as in [6] and differ slightly in form from those of [4, 5].

It is worthy of remark that all information in Eqs. (3.8) and (3.9) about the intermolecular potential $K(r)$, the pair correlation function $g(r)$ and even about the crystal periods $\mathbf{d}_1$, $\mathbf{d}_2$ and $\mathbf{d}_3$ resides only in $\sigma(A)$. The quantity $\sigma(0)$ that plays an important role for fluids [6] is not present in (3.8) and (3.9). It is not even necessary to know the full function $\sigma(k)$ for Eqs. (3.8) and (3.9) contain the value of the function only at a discrete set of points in reciprocal space since $A = |l\mathbf{a}_1 + m\mathbf{a}_2 + n\mathbf{a}_3|$. Aside from $\sigma(A)$, only two external parameters $\rho_0$ and $\theta$ figure in (3.8) and (3.9). In this paper we shall neglect the dependence of $\sigma(A)$ upon the temperature $\theta$ because the dependence should be weak for crystals according to Section 4 of Ref. [6] and does not tell upon principal results obtained in the present paper. If necessary, the dependence can be taken into account when dealing with a concrete situation.

For what follows we need also the Helmholtz free energy (Eq. (5.9) of [6])

$$F = F_f - N\theta \ln G - \frac{V}{2} \sum_{l,m,n}{}' |a_{lmn}|^2 \sigma(A), \qquad (3.10)$$

in which $F_f$ is the free energy of the corresponding fluid that is not required for our purposes.

Ref. [5] demonstrates that the concrete form of the Fourier series of (3.5) is different for different space groups and, moreover, is characteristic of each space group (only cubic space groups were considered in that reference). The form of the Fourier series specifies a definite form of the system of equations for the Fourier coefficients $a_{lmn}$, which in its turn determines peculiarities of solutions to the set and eventually influences properties of the relevant crystal and phase transitions in it.

## 4. Parent phase

We proceed now to a study of the high-temperature parent phase with space group $D_{2h}^{16}$ from which the phase transitions considered in the following sections originate. As long as we shall utilize tables of Ref. [15] in the next sections, we adopt the description of space groups given in [15], which implies an appropriate choice of the coordinate system. First of all, we should establish the form of the Fourier series if space group $D_{2h}^{16}$ is involved.

The idea as to how to find out the form of the Fourier series is outlined in Appendix of [5]. Here we shall consider this in more detail. Seeing that space group $D_{2h}^{16}$ pertains to the orthorhombic system, instead of the general form of (3.5) the Fourier series can be written in the form

$$\rho(\mathbf{r}) = \sum_{l,m,n=-\infty}^{\infty} a_{lmn}\, e^{i(la_1 x + ma_2 y + na_3 z)}. \tag{4.1}$$

We apply each of 8 symmetry transformations of space group $D_{2h}^{16}$ to the series. The transformations consist in changing the signs of $x$, $y$, $z$ [15] and in translations by half-periods that add $\pi$ in the bracket in the exponential since the periods are $d_1 = 2\pi/a_1$, $d_2 = 2\pi/a_2$, $d_3 = 2\pi/a_3$. After the transformation, we change the signs of the variables of summation $l$, $m$, $n$ in order to have the same factor $\exp i(la_1 x + ma_2 y + na_3 z)$ and equate the coefficients of the factor in the initial series and in the series obtained. As a result, we have

$$a_{lmn} = a_{l,-m,-n}(-1)^{l+m} = a_{-l,m,-n}(-1)^{m+n} = a_{-l,-m,n}(-1)^{l+n} = a_{-l,-m,-n}(-1)^{l+m}$$

$$= a_{-l,m,n} = a_{l,-m,n}(-1)^{l+n} = a_{l,m,-n}(-1)^{m+n}. \tag{4.2}$$

This together with the condition $a_{lmn}^{*} = a_{-l-m-n}$ which expresses the fact that $\rho(\mathbf{r})$ is real-valued determines the form of the Fourier coefficients in the present case. For example, all $a_{lmn}$ are pure real or pure imaginary; $a_{l00} \neq 0$, $a_{0m0} \neq 0$, $a_{00n} \neq 0$ only if the integers $l$, $m$, $n$ are even.

In our case it is convenient to consider the effective potential $U(\mathbf{r})$ of (3.6) that has the same symmetry as $\rho(\mathbf{r})$ seeing that $\sigma(A) = \sigma\left(\sqrt{l^2 a_1^2 + m^2 a_2^2 + n^2 a_3^2}\right)$ does not change the symmetry (this follows from (3.1) as well). As a result, first terms in $U(\mathbf{r})$ for $D_{2h}^{16}$ are of the form

$$U(\mathbf{r}) = \rho_0 \sigma(0) + 4\alpha_1 \sigma\left(\sqrt{a_1^2 + a_3^2}\right) \cos a_1 x \, \sin a_3 z + 4\alpha_2 \sigma\left(\sqrt{a_2^2 + a_3^2}\right) \sin a_2 y \, \cos a_3 z$$

$$+ 8\alpha_3' \sigma\left(\sqrt{a_1^2 + a_2^2 + a_3^2}\right) \cos a_1 x \, \cos a_2 y \, \cos a_3 z$$

$$+ 2\alpha_4' \sigma(2a_1)\cos 2a_1 x + 2\alpha_4'' \sigma(2a_2)\cos 2a_2 y + 2\alpha_4''' \sigma(2a_3)\cos 2a_3 z + \ldots, \tag{4.3}$$

where

$$\alpha_1 = ia_{101}, \ \alpha_2 = ia_{011}, \ \alpha_3' = a_{111}, \ \alpha_4' = a_{200}, \ \alpha_4'' = a_{020}, \ \alpha_4''' = a_{002}. \tag{4.4}$$

For simplicity's sake, in the following we shall assume that $a_1 = a_2 \approx a_3$ (it is seen from (4.3) that the behaviour of $U(\mathbf{r})$ in the $z$-direction differs from its behaviour in the $x$- and $y$-directions).

In the following we shall use the Kirkwood approximation for a crystal [16, 5]. The approximation consists in putting $\sigma(A) = 0$ if $A > A_0$ in the effective potential $U(\mathbf{r})$ of (3.6) and in other formulae. In Ref. [6] it is demonstrated using the Lennard-Jones potential as an example that the Kirkwood approximation is rather good for the crystal. A systematic method for improving the approximation is discussed in the concluding section of Ref. [5]. It should be underlined that, although one retains only several Fourier harmonics in $U(\mathbf{r})$, the density $\rho(\mathbf{r})$ contains all harmonics in view of (3.1). Moreover, the Fourier coefficients $a_{lmn}$ of the density may be large as long as the Kirkwood approximation does not concern $a_{lmn}$ in (3.6) (see examples in [5]). It is to be added that the quantities $\sigma(A)$ even with small $A$ which are retained depend upon the whole intermolecular potential $K(r)$ including its short-range part according to (3.7). Besides, the function $K_g(r)$ that figures in (3.7) should be rather smooth. This is seen from (3.4) because, even if $K(r') \to \infty$ as $r' \to 0$, then $g(r') \to 0$. Extra smoothing is due to the integration over $r'$ in (3.4). Owing to the smoothness of $K_g(r)$ the Fourier harmonics of $U(\mathbf{r})$ are small for large $A$ even in the case of the short-range and highly singular Lennard-Jones potential mentioned above.

Implying the Kirkwood approximation we suppose that $\sigma(\zeta a_i) = 0$ if $\zeta > \sqrt{2}$, in which case Eq. (4.3) becomes

$$U(\mathbf{r}) = \rho_0\sigma(0) + 4\alpha_1\sigma_1\cos a_1 x \sin a_3 z + 4\alpha_2\sigma_1\sin a_2 y \cos a_3 z, \qquad (4.5)$$

where $\sigma_1 = \sigma\left(\sqrt{a_1^2 + a_3^2}\right) = \sigma\left(\sqrt{a_2^2 + a_3^2}\right)$.

We see from (4.5) that with the approximation used it is sufficient to deduce equations for $\alpha_1$ and $\alpha_2$ alone. This can be done with the help of (3.8) and (4.4). When reducing Eq. (3.8) it is necessary to prove that some integrals vanish. This can be achieved by appropriate replacement of the variables of integration where $U(\mathbf{r})$ remains invariant and exploiting the following fact. If one has a periodic function $f(x)$, that is to say, if $f(x + d) = f(x)$, then

$$\int_0^d f(\pm x + c)dx = \int_0^d f(x)dx \qquad (4.6)$$

with an arbitrary constant $c$. As a result, we shall obtain the following set of transcendental equations

$$\frac{\alpha_1}{\rho_0} = \frac{I_1^{(1)}(\beta_1,\beta_2)}{I_0^{(1)}(\beta_1,\beta_2)}, \qquad \frac{\alpha_2}{\rho_0} = \frac{I_2^{(1)}I(\beta_1,\beta_2)}{I_0^{(1)}(\beta_1,\beta_2)}, \qquad (4.7)$$

where

$$I_0^{(1)}(\beta_1,\beta_2) = \frac{1}{8\pi^3}\int_0^{2\pi}\int_0^{2\pi}\int_0^{2\pi} e^{-4(\beta_1\cos\xi_1\sin\xi_3 + \beta_2\sin\xi_2\cos\xi_3)}d\xi_1 d\xi_2 d\xi_3, \qquad (4.8)$$

$$I_1^{(1)}(\beta_1,\beta_2) = \frac{1}{8\pi^3} \int_0^{2\pi}\int_0^{2\pi}\int_0^{2\pi} \cos\xi_1 \sin\xi_3 e^{-4(\beta_1\cos\xi_1\sin\xi_3 + \beta_2\sin\xi_2\cos\xi_3)} d\xi_1 d\xi_2 d\xi_3, \qquad (4.9)$$

$$I_2^{(1)}(\beta_1,\beta_2) = \frac{1}{8\pi^3} \int_0^{2\pi}\int_0^{2\pi}\int_0^{2\pi} \sin\xi_2 \cos\xi_3 e^{-4(\beta_1\cos\xi_1\sin\xi_3 + \beta_2\sin\xi_2\cos\xi_3)} d\xi_1 d\xi_2 d\xi_3; \qquad (4.10)$$

$$\beta_1 = \frac{\alpha_1\sigma_1}{\theta}, \qquad \beta_2 = \frac{\alpha_2\sigma_1}{\theta}. \qquad (4.11)$$

It can be readily shown that

$$I_1^{(1)}(\beta_1,\beta_2) = -\frac{1}{4}\frac{\partial I_0^{(1)}(\beta_1,\beta_2)}{\partial\beta_1}, \quad I_2^{(1)}(\beta_1,\beta_2) = -\frac{1}{4}\frac{\partial I_0^{(1)}(\beta_1,\beta_2)}{\partial\beta_2}, \quad I_1^{(1)}(\beta_1,\beta_2) = I_2^{(1)}(\beta_2,\beta_1). \quad (4.12)$$

Consequently it is sufficient to investigate the integral $I_0^{(1)}$ alone. The simplest method for calculating $I_0^{(1)}$ is to expand the exponential of (4.8) in a series with the result (cf. [5])

$$I_0^{(1)}(\beta_1,\beta_2) = I_0^{(1)}(\beta_2,\beta_1) = \sum_{n=0}^{\infty}\frac{1}{n!}\sum_{m=0}^{n}\frac{(2n-2m)!(2m)!}{(n-m)!^3 (m)!^3}\beta_1^{2m}\beta_2^{2n-2m}. \qquad (4.13)$$

With use made of (4.7) the equations of (4.11) can be rewritten as

$$\frac{\theta}{\rho_0\sigma_1} = \frac{I_1^{(1)}(\beta_1,\beta_2)}{\beta_1 I_0^{(1)}(\beta_1,\beta_2)}, \qquad \frac{\beta_2 I_1^{(1)}(\beta_1,\beta_2)}{\beta_1 I_2^{(1)}(\beta_1,\beta_2)} = 1. \qquad (4.14)$$

In actual fact, the equations of (4.7) and (4.14) directly yield the temperature dependence of $\alpha_1$ and $\alpha_2$ in a parametric form, $\beta_1$ and $\beta_2$ being the parameters. At a given temperature $\theta$, from the two equations of (4.14) we find $\beta_1$ and $\beta_2$ which give the corresponding values of $\alpha_1$ and $\alpha_2$ when substituted into (4.7). According to (4.8), (4.13) and (4.12), always $I_0^{(1)} > 0$ whereas $I_1^{(1)}$ is opposite in sign to $\beta_1$. Therefore the right-hand side of the first equation of (4.14) is always negative and in consequence the solution exists only if $\sigma_1 < 0$ (cf. [5]).

An analysis shows that there exist three nontrivial solutions

$$(1)\ \alpha_1 \neq 0,\ \alpha_2 = 0, \quad (2)\ \alpha_1 = 0,\ \alpha_2 \neq 0, \quad (3)\ \alpha_1 = \alpha_2. \qquad (4.15)$$

The solutions make their appearance when $\beta_1$ and $\beta_2$ are small, and the condition for this can be found upon placing the expansion of (4.13) in (4.7). All three solutions exist if $\theta \leq -\rho_0\sigma_1$. At $\theta = -\rho_0\sigma_1 = \rho_0|\sigma_1|$ there occurs a phase transition from the liquid to a $D_{2h}^{16}$ crystal.

Now, by (3.10), we can calculate the difference $F - F_f$ of the Helmholtz free energies for the crystal and liquid. With the same assumption for $\sigma(A)$ as in (4.5), in dimensionless units we have

$$F_c \equiv \frac{F - F_f}{\rho_0 / \sigma_1 / N} = 2\frac{\alpha_1^2 + \alpha_2^2}{\rho_0^2} - \tilde{\theta}\ln I_0^{(1)}, \qquad (4.16)$$

where $\tilde{\theta} = \theta / \rho_0 / \sigma_1 /$ is a dimensionless temperature.

The result of numerical calculations for the temperature dependence of $F_c$ is presented in Fig. 2. The figure shows that the liquid-crystal phase transition should be second order in the present case. An argument was adduced in [5] that the second-order phase transition between a fluid and crystal is in fact impossible. This is of no importance for the present investigation inasmuch as we are interested in the $D_{2h}^{16}$ crystal itself and not in the question as to how it can be obtained. It is to be added that we have set $a_1 = a_2$ above (if $a_1 \neq a_2$, Fig. 2 would be more complicated) and have used the simplified form of $U(\mathbf{r})$ of (4.5) instead of (4.3) in order to have the simplest variant of the $D_{2h}^{16}$ phase because the phase plays an auxiliary role for the studies in this paper.

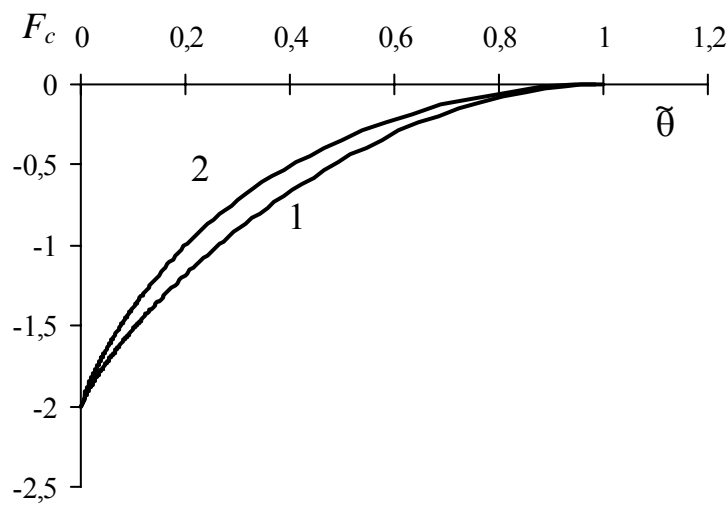

**Fig. 2**. Temperature dependence of $F_c$ defined in (4.16) for space group $D_{2h}^{16}$.

Let us point out nevertheless some curious results that follow from Fig. 2. Curve 1 corresponds to solutions 1 or 2 of (4.15) while curve 2 to solution 3. Factually, solutions 1 and 2 lead to two-dimensional crystals because, if $\alpha_2 = 0$, it is seen from (4.5) that $U(\mathbf{r})$ and consequently $\rho(\mathbf{r})$ do not depend on $y$ whereas, if $\alpha_1 = 0$, $U(\mathbf{r})$ and $\rho(\mathbf{r})$ do not depend on $x$. Therefore, Fig. 2 indicates that, in the case under consideration, the two-dimensional crystals are energetically more preferable than the three-dimensional one relevant to solution 3. This is due to the fact that we used the simplified form of $U(\mathbf{r})$ of (4.5) instead of (4.3). Besides, as noted in Introduction a $D_{2h}^{16}$ crystal composed of particles of one kind with spherically symmetric interaction can be in a metastable state alone. In the following sections, we shall study phase transitions implying that the parent phase corresponds to solution 3 of (4.15) although the phase is metastable.

## 5. Commensurate phase transition

Seeing that our studies are based upon the sequence of phase transitions that occur in ammonium fluoberyllate, let us recall the relevant phase transitions [17,18]. The high-temperature $D_{2h}^{16}$ phase exists at $T > -90°$ C, between $-90°$ C and $-96°$ C an incommensurate phase is observed, and below $T = -96°$ C there emerges a polar commensurate phase with space group $C_{2v}^9$ in which the polarization is directed along the $z$-axis and the period along the $y$-axis doubles (we choose the coordinate system as in Section 4; another coordinate system is usually taken in experimental papers: the polar axis is the $y$-axis and the $x$-period doubles).

As is done in the Landau theory, we should first consider the transition from a $D_{2h}^{16}$ phase to a $C_{2v}^9$ phase. When analysing the transition in the Landau theory one finds that there exists a Lifshitz invariant and therefore a direct transition from the $D_{2h}^{16}$ phase to the $C_{2v}^9$ phase is impossible (see Section 2). In the present section, we shall ignore the existence of the Lifshitz invariant and, using this example, show how a commensurate phase transition can be treated in statistical theory. This procedure corresponds to obtaining the solutions of (2.3) or (2.4).

First of all, it needs to establish the form of the Fourier series in the case of space group $C_{2v}^9$. Space group $C_{2v}^9$ pertains to the same orthorhombic system, and we have again (4.1). We must take a coordinate system compatible with the one chosen in the previous section for space group $D_{2h}^{16}$. To this end, the $x$- and $y$-axes adopted in [15] for space group $C_{2v}^9$ are to be interchanged. Now, analogously with (4.2), in the present instance one obtains the following conditions on the Fourier coefficients:

$$a_{lmn} = a_{-l,-m,n}(-1)^{l+m+n} = a_{-l,m,n}(-1)^m = a_{l,-m,n}(-1)^{l+n}. \tag{5.1}$$

With the help of these conditions we find the potential $U(\mathbf{r})$ for space group $C_{2v}^9$:

$$U(\mathbf{r}) = \rho_0\sigma(0) + 4\alpha'_2\sigma\!\left(\sqrt{a_1^2 + a_3^2}\right)\cos a_1 x \cos a_3 z + 4\alpha''_2\sigma\!\left(\sqrt{a_1^2 + a_3^2}\right)\cos a_1 x \sin a_3 z$$
$$+ 4\alpha'''_2\sigma\!\left(\sqrt{a_1^2 + a_2^2}\right)\sin a_1 x \sin a_2 y + 8\alpha'_3\sigma\!\left(\sqrt{a_1^2 + a_2^2 + a_3^2}\right)\sin a_1 x \cos a_2 y \sin a_3 z$$
$$+ 8\alpha''_3\sigma\!\left(\sqrt{a_1^2 + a_2^2 + a_3^2}\right)\sin a_1 x \cos a_2 y \cos a_3 z + 2\alpha'_4\sigma(2a_1)\cos 2a_1 x + 2\alpha''_4\sigma(2a_2)\cos 2a_2 y$$
$$+ 2\alpha'''_4\sigma(2a_3)\cos 2a_3 z + 2\alpha'''_4\sigma(2a_3)\sin 2a_3 z + \ldots, \tag{5.2}$$

where

$$\alpha'_2 - i\alpha''_2 = a_{101},\ \alpha'''_2 = -a_{110},\ \alpha'_2 + i\alpha''_2 = -a_{111},\ \alpha'_4 = a_{200},\ \alpha''_4 = a_{020},\ \alpha'''_4 - i\alpha''''_4 = a_{002}. \tag{5.3}$$

At this point it is worth remarking that another method for writing down the Fourier series for a given space group could be utilized. When applying symmetry operations to the series of (4.1) one can see that the quantities $a_1x$, $a_2y$, and $a_3z$ transform in a definite way. These transformations for space group $C_{2v}^9$ are listed in Table 1 where the symbols $h_i$ of [15] are used for operations. One may readily verify that all terms in (5.2) remain invariant under the transformations of Table 1. Making use of the table one can write down any desired term in the Fourier series for $U(\mathbf{r})$ and $\rho(\mathbf{r})$. A similar table can be compiled for space group $D_{2h}^{16}$.

**Table 1**. Symmetry transformations for space group $C_{2v}^9$.

| Initial quantity | Transformed quantity | | |
|---|---|---|---|
| | $h_4$ | $h_{26}$ | $h_{27}$ |
| $a_1x$ | $-a_1x + \pi$ | $-a_1x$ | $a_1x + \pi$ |
| $a_2y$ | $-a_2y + \pi$ | $a_2y + \pi$ | $-a_2y$ |
| $a_3z$ | $a_3z + \pi$ | $a_3z$ | $a_3z + \pi$ |

Proceeding in line with the Landau theory we should now identify the irreducible representation of space group $D_{2h}^{16}$ responsible for the $D_{2h}^{16} - C_{2v}^9$ transition. Seeing that the period in the $C_{2v}^9$ phase doubles along the $y$-axis, the representation must correspond to the point $K = a_2/2$ of the Brillouin zone (see also Section 2). According to Kovalev's tables [15] there are two irreducible representations $T_1$ and $T_2$ under number $T70$ relevant to this point. An analysis similar to the one carried out below for $T_2$ indicates that $T_1$ does not lead to a $C_{2v}^9$ phase, so that we shall consider only $T_2$.

In the Landau theory [19], in the case of a two-component order parameter the density below the transition point acquires the form

$$\rho(\mathbf{r}) = \rho_0(\mathbf{r}) + C_1\varphi_1(\mathbf{r}) + C_2\varphi_2(\mathbf{r}), \tag{5.4}$$

in which $\rho_0(\mathbf{r})$ is invariant under the symmetry operations of the parent phase while $\varphi_1(\mathbf{r})$ and $\varphi_2(\mathbf{r})$ transform according to the representation $T_2$. If one applies a symmetry operation of the parent phase to (5.4), one obtains

$$\rho(\mathbf{r}) = \rho_0(\mathbf{r}) + (C_1T_{11} + C_2T_{12})\varphi_1(\mathbf{r}) + (C_1T_{21} + C_2T_{22})\varphi_2(\mathbf{r}), \tag{5.5}$$

where $T_{ij}$ are the matrix elements of $T_2$. The density $\rho(\mathbf{r})$ remains unchanged if (5.4) and (5.5) coincide, which is possible only for some symmetry operations. The matrix elements of $T_2$ when converted to a real-valued form are 0 or $\pm 1$ alone and therefore it is sufficient to consider three cases: (1) $C_1 = \pm C_2$, (2) $C_2 = 0$ or $C_1 = 0$, (3) all other cases. Analysing the symmetry operations that conserve the form of ρ(**r**) one can find out the relevant space group (sometimes an appropriate change of the coordinate system is required when comparing the results with a list of space group). The analysis shows that space group $C_{2v}^9$ results in the first case, space group $C_{2h}^5$ in the second one, and space group $C_s^2$ in all other cases (it may be noted in passing that the above representation $T_1$ leads to space groups $C_{2v}^7$, $C_{2h}^2$ and $C_s^1$). In the Landau theory, it remains only to determine the number of different invariants composed of φ$_1$ and φ$_2$ (replaced further by η and ξ), which can be done with use made of general methods [20]; in particular, one will see that there is a Lifshitz invariant in this case. Thereupon one can at once write down the free energy of (2.1) as the form of the invariants may be readily established. It should be observed that the above symmetry analysis yields three commensurate phases whereas in Section 2 we had only two phases according to (2.3) and (2.4). The solution where $0 \neq \eta \neq \xi \neq 0$ describing the third commensurate phase appears if invariants of sixth order in η and ξ are added in (2.1).

In the statistical approach it needs to know the basis functions φ$_1$(**r**) and φ$_2$(**r**) of (5.4). It is not difficult to surmise a possible form of the functions: φ$_1$ = $\sin a_1 x \exp(-i a_2 y/2)$ and φ$_2$ = $\sin a_1 x \exp(i a_2 y/2)$, and to verify that they transform according to the representation $T_2$. For our purposes we take linear combinations of these functions that are real-valued:

$$\varphi_1' = \sin a_1 x \sin \frac{a_2 y}{2}, \qquad \varphi_2' = \sin a_1 x \cos \frac{a_2 y}{2}. \tag{5.6}$$

It should be remarked that the linear transformation changes the matrix elements $T_{ij}$ and now space group $C_{2v}^9$ corresponds to the cases where $C_1 \neq 0$, $C_2 = 0$ or vice versa. These two cases are in fact identical because the functions of (5.6) go over into each other if the coordinate origin is shifted along the y-axis, and we shall imply the case where $C_1 \neq 0$, $C_2 = 0$, which is in agreement with the coordinate system chosen in (5.2).

In the statistical approach, an equation of the type (5.4) should be written for the potential $U(\mathbf{r})$ rather than for ρ(**r**) because the density ρ(**r**) becomes δ-like as θ → 0 while the functions of (5.6) are smooth. Thus below the phase transition point, when a wave corresponding to $\varphi_1'$ is frozen-in in the crystal (cf. Section 2), the potential $U(\mathbf{r})$ acquires the form

$$U(\mathbf{r}) = U_0(\mathbf{r}) + C_1 \sin a_1 x \sin \frac{a_2 y}{2}, \tag{5.7}$$

where $U_0(\mathbf{r})$ is of $D_{2h}^{16}$ symmetry. We take $U(\mathbf{r})$ of (4.5) for $U_0(\mathbf{r})$, which yields the potential that will enable us to study the phase transition and properties of the emerging $C_{2v}^9$ phase:

$$U(\mathbf{r}) = \rho_0\sigma(0) + 4\alpha_1\sigma_1\cos a_1x \sin a_3z + 4\alpha_2\sigma_1\sin 2a_2'y \cos a_3z + 4\alpha_3\sigma_3\sin a_1x \sin a_2'y, \quad (5.8)$$

where $a_2' = a_2/2$ and $\sigma_3 = \sigma\left(\sqrt{a_1^2 + a_2'^2}\right) = \sigma\left(\sqrt{a_1^2 + a_2^2/4}\right)$.

If one compares (5.8) with (5.2), one sees that (5.8) is a special case of (5.2). The terms of (5.8) with $\cos a_1x \sin a_3z$ and $\sin a_1x \sin a_2'y$ are present in (5.2) (the comparison yields the unknown coefficient $C_1$ of (5.7)). Although the term of (5.8) with $\sin 2a_2'y \cos a_3z$ is not written down in (5.2), it is admitted by the $C_{2v}^9$ symmetry, which may be checked with the aid of Table 1. Only the term with $\cos a_1x \cos a_3z$ which figures in the general series of (5.2) and which is admitted by the Kirkwood approximation used for (4.5) is missing from (5.8). This endows $U(\mathbf{r})$ of (5.8) with extra symmetry, namely, the potential remains unchanged under the transformation

$$a_1x \to -a_1x + \pi, \quad a_3z \to -a_3z. \quad (5.9)$$

Strictly speaking, the Kirkwood approximation allows of one more term, namely, the term with $\cos 2a_2'y$ from (5.2). For simplicity's sake we discard this term implying that $\sigma(2a_2') = \sigma(a_2) \approx 0$.

We should next deduce equations for the parameters $\alpha_1$, $\alpha_2$ and $\alpha_3$ that enter into (5.8), which can be done with use made of Eq. (3.8) as in Section 4. In place of (4.4) and (5.3) we have now $\alpha_1 = ia_{101}$, $\alpha_2 = ia_{021}$, $\alpha_3 = -a_{110}$. In this connection the following is worthy of remark. According to (5.3) the Fourier coefficient $a_{101}$ contains a real part, namely, $\alpha_2'$. However, when calculating $a_{101}$ with use made of (3.8) one will see that the real part disappears owing to (5.9). As a result, analogously to (4.7)–(4.10) we shall obtain that

$$\frac{\alpha_1}{\rho_0} = \frac{I_1^{(2)}(\beta_1,\beta_2,\beta_3)}{I_0^{(2)}(\beta_1,\beta_2,\beta_3)}, \quad \frac{\alpha_2}{\rho_0} = \frac{I_2^{(2)}(\beta_1,\beta_2,\beta_3)}{I_0^{(2)}(\beta_1,\beta_2,\beta_3)}, \quad \frac{\alpha_3}{\rho_0} = \frac{I_3^{(2)}(\beta_1,\beta_2,\beta_3)}{I_0^{(2)}(\beta_1,\beta_2,\beta_3)}, \quad (5.10)$$

where

$$I_0^{(2)}(\beta_1,\beta_2,\beta_3) = \frac{1}{8\pi^3}\int_0^{2\pi}\int_0^{2\pi}\int_0^{2\pi} e^{-4(\beta_1\cos\xi_1\sin\xi_3 + \beta_2\sin 2\xi_2\cos\xi_3 + \beta_3\sin\xi_1\sin\xi_2)}d\xi_1 d\xi_2 d\xi_3, \quad (5.11)$$

$$I_1^{(2)}(\beta_1,\beta_2,\beta_3) = \frac{1}{8\pi^3}\int_0^{2\pi}\int_0^{2\pi}\int_0^{2\pi} \cos\xi_1\sin\xi_3 e^{-4(\beta_1\cos\xi_1\sin\xi_3 + \beta_2\sin 2\xi_2\cos\xi_3 + \beta_3\sin\xi_1\sin\xi_2)}d\xi_1 d\xi_2 d\xi_3, \quad (5.12)$$

$$I_2^{(2)}(\beta_1,\beta_2,\beta_3) = \frac{1}{8\pi^3}\int_0^{2\pi}\int_0^{2\pi}\int_0^{2\pi} \sin 2\xi_2\cos\xi_3 e^{-4(\beta_1\cos\xi_1\sin\xi_3 + \beta_2\sin 2\xi_2\cos\xi_3 + \beta_3\sin\xi_1\sin\xi_2)}d\xi_1 d\xi_2 d\xi_3, \quad (5.13)$$

$$I_3^{(2)}(\beta_1,\beta_2,\beta_3) = \frac{1}{8\pi^3}\int_0^{2\pi}\int_0^{2\pi}\int_0^{2\pi} \sin\xi_1\sin\xi_2 e^{-4(\beta_1\cos\xi_1\sin\xi_3 + \beta_2\sin 2\xi_2\cos\xi_3 + \beta_3\sin\xi_1\sin\xi_2)}d\xi_1 d\xi_2 d\xi_3, \quad (5.14)$$

$$\beta_1 = \frac{\alpha_1\sigma_1}{\theta}, \quad \beta_2 = \frac{\alpha_2\sigma_1}{\theta}, \quad \beta_3 = \frac{\alpha_3\sigma_3}{\theta}. \quad (5.15)$$

From these formulae it follows that

$$I_1^{(2)} = -\frac{1}{4}\frac{\partial I_0^{(2)}}{\partial \beta_1}, \quad I_2^{(2)} = -\frac{1}{4}\frac{\partial I_0^{(2)}}{\partial \beta_2}, \quad I_3^{(2)} = -\frac{1}{4}\frac{\partial I_0^{(2)}}{\partial \beta_3}. \tag{5.16}$$

The integral $I_0^{(2)}$ can be calculated with the help of the series

$$I_0^{(2)}(\beta_1,\beta_2,\beta_3) = \sum_{n=0}^{\infty}\sum_{m=0}^{n}\frac{\beta_3^{2m}}{m!(n-m)!}\sum_{l=0}^{n-m}\frac{(2n-2l)!(2n-2m-2l)!(2l)!\beta_1^{2l}\beta_2^{2n-2m-2l}}{(l!)^2(n-m-l)!^2(n-l)!(m+l)!(2n-m-2l)!}. \tag{5.17}$$

With use made of (5.10) the equations of (5.15) can be recast as

$$\frac{\theta}{\rho_0\sigma_1} = \frac{I_1^{(2)}(\beta_1,\beta_2,\beta_3)}{\beta_1 I_0^{(2)}(\beta_1,\beta_2,\beta_3)}, \quad \frac{\beta_2 I_1^{(2)}(\beta_1,\beta_2,\beta_3)}{\beta_1 I_2^{(2)}(\beta_1,\beta_2,\beta_3)} = 1, \quad \frac{\beta_3 I_1^{(2)}(\beta_1,\beta_2,\beta_3)}{\beta_1 I_3^{(2)}(\beta_1,\beta_2,\beta_3)} = \frac{\sigma_3}{\sigma_1}. \tag{5.18}$$

These equations together with Eq. (5.10) yield the temperature dependence of $\alpha_1$, $\alpha_2$ and $\alpha_3$ with $\beta_1$, $\beta_2$, $\beta_3$ as parameters. On the base of (3.10), by analogy with (4.16), we calculate the Helmholtz free energy $F_c$ relevant to the crystal

$$F_c \equiv \frac{F-F_f}{\rho_0/\sigma_1/N} = 2\frac{\alpha_1^2+\alpha_2^2+\sigma_3\alpha_3^2/\sigma_1}{\rho_0^2} - \tilde{\theta}\ln I_0^{(2)} \tag{5.19}$$

with the same dimensionless temperature $\tilde{\theta} = \theta/\rho_0/\sigma_1/$.

If $\beta_3 = 0$, one can see that $I_0^{(2)} = I_0^{(1)}$, $I_1^{(2)} = I_1^{(1)}$, $I_2^{(2)} = I_2^{(1)}$ and $I_3^{(2)} = 0$, so that one has the parent phase of Section 4. If $\beta_3 \neq 0$, one has the $C_{2v}^9$ phase. Passing to the limit as $\beta_3 \to 0$ with $\beta_1 = \beta_2$ in the last equation of (5.18) one arrives at the condition for formation of the $C_{2v}^9$ phase:

$$\frac{\sigma_3}{\sigma_1} = -\frac{I_1^{(1)}(\beta_1,\beta_1)}{4\beta_1 I_{31}^{(2)}(\beta_1)}, \tag{5.20}$$

where

$$I_{31}^{(2)}(\beta_1) = \frac{1}{8\pi^3}\int_0^{2\pi}\int_0^{2\pi}\int_0^{2\pi}\sin^2\xi_1\sin^2\xi_2 e^{-4\beta_1(\cos\xi_1\sin\xi_3+\sin 2\xi_2\cos\xi_3)}d\xi_1 d\xi_2 d\xi_3$$

$$= \frac{1}{16\pi^3}\int_0^{2\pi}\int_0^{2\pi}\int_0^{2\pi}\sin^2\xi_1 e^{-4\beta_1(\cos\xi_1\sin\xi_3+\sin\xi_2\cos\xi_3)}d\xi_1 d\xi_2 d\xi_3. \tag{5.21}$$

The right-hand side of Eq. (5.20) is positive (see the remark in Section 4 as to the sign of $I_1^{(1)}$) and thereby $\sigma_3$ must be negative since $\sigma_1 < 0$. Numerical calculation shows that this right-hand side tends to unity as $\beta_1 \to 0$ or $\beta_1 \to \infty$, and has a minimum equal to 0.7890 at $\beta_1 = 1.6089$. Hence, the $C_{2v}^9$ phase under study can exist only if $0.7890 < \sigma_3/\sigma_1 < 1$. Eq. (5.20) admits two solutions for each value of $\sigma_3/\sigma_1$ from this interval. The temperature at which the phase transition to the $C_{2v}^9$ phase occurs can be found from the first equation of (5.18) in the same limit as $\beta_3 \to 0$ with $\beta_1 = \beta_2$ found from (5.20). With use made of (5.20) the relevant dimensionless temperature can be conveniently written as

$$\tilde{\theta} \equiv \frac{\theta}{\rho_0|\sigma_1|} = 4\frac{\sigma_3 I_{31}^{(2)}(\beta_1)}{\sigma_1 I_0^{(1)}(\beta_1, \beta_1)}. \tag{5.22}$$

The behaviour of $\alpha_1$, $\alpha_2$, $\alpha_3$ and of the free energy $F_c$ of (5.19) near the phase transition point can be elucidated by expanding the integrals of (5.11)–(5.14) in powers of $\beta_3$ with account taken of the fact that $\beta_1 \ne \beta_2$ in the emerging phase although $\beta_1 = \beta_2$ in the parent phase. The relevant procedure is similar, though more involved, to the one employed for deducing Eqs. (6.13) and (6.14) in [5]. Here we shall not describe the procedure and shall not adduce the resulting cumbersome formulae which are of only academic interest by limiting ourselves to the results of numerical calculations for the entire temperature region where the $C_{2v}^9$ phase exists. The calculations were performed for $\sigma_3/\sigma_1 = 0.9$. The second-order phase transition into the $C_{2v}^9$ phase happens at $\tilde{\theta} = 0.726$; as the temperature lowers, the value of $\beta_3$ increases from 0 to 0.522 and thereafter begins to decrease and eventually vanishes at $\tilde{\theta} = 0.149$ where an inverse second-order transition from the $C_{2v}^9$ phase to the $D_{2h}^{16}$ phase occurs. We do not represent the curve describing the $C_{2v}^9$ phase in Fig. 2 as long as the curve, passing slightly below curve 2, will practically merge with the last at the scale of the figure. The existence of the inverse phase transition may be due to the peculiarities of space group $D_{2h}^{16}$ for the intermolecular interactions implied in this paper as pointed out in Introduction.

Closing the section it is worthy of remark that the role of the order parameter in the present case is played by $\beta_3$, more precisely by $\alpha_3 = -a_{110}$, i.e., by the Fourier coefficient $a_{110}$. A similar situation holds for the phase transition considered in [5] (see the discussion of Eq. (6.16) in [5]).

## 6. Incommensurate phase

First of all, it needs to establish the form of the density $\rho(\mathbf{r})$ for the incommensurate phase. In view of the remark at the end of the preceding section the role of the order parameter in the present approach is played by Fourier coefficients $a_{lmn}$. According to the Landau theory (Section 2) the order parameter in the incommensurate phase is spatially modulated. Consequently, the coefficients $a_{lmn}$ in the incommensurate phase should be of the form

$$a_{lmn} = \sum_{p=-\infty}^{\infty} a_{lmnp}\, e^{ipka_2' y}, \tag{6.1}$$

where, for convenience, we have introduced $a'_2$ into the exponent in order that the parameter $k$ be dimensionless. Being irrational the parameter characterizes the periodic modulation of $a_{lmn}$ in the incommensurate phase.

Substituting (6.1) into (4.1) with $a_2 \to a'_2$ (we should have the $C^9_{2v}$ phase if $k = 0$) yields

$$\rho(\mathbf{r}) = \sum_{l,m,n,p=-\infty}^{\infty} a_{lmnp}\, e^{i(la_1 x + ma'_2 y + na_3 z + pka'_2 y)} . \tag{6.2}$$

This nonperiodical three-dimensional density $\rho(\mathbf{r})$ is characterized by four parameters $a_1$, $a'_2$, $a_3$ and $ka'_2$. It is worth remarking that the symmetry of such a density may be described with the aid of superspace groups [21, 22]. By Eq. (3.3) analogously with (3.6), we can also calculate the potential $U(\mathbf{r})$ if $\rho(\mathbf{r})$ is given by (6.2):

$$U(\mathbf{r}) = \sum_{l,m,n,p=-\infty}^{\infty} a_{lmnp}\, \sigma\!\left(\sqrt{l^2 a_1^2 + (m+kp)^2 a_1'^2 + n^2 a_3^2}\right) e^{i(la_1 x + ma'_2 y + na_3 z + pka'_2 y)} . \tag{6.3}$$

The present approach leans heavily on the symmetry of Fourier series in the case of space groups. At the same time, the series of (6.2) is not a Fourier series if $k$ is irrational. For this reason we adopt the following strategy. An irrational number $k$ can be represented with any desired precision by a ratio of two integers $\mu$ and $\nu$, namely, $k = \mu/\nu$. If, for example, one wants to have $k$ with an accuracy of $n$ digits after the decimal point, it is sufficient to set $\nu = 10^n$, to multiply $k$ by $10^n$ and to discard all digits after the decimal point in the resulting number, which will give the integer $\mu$. Therefore, the irrational number $k$ can be represented as $k = \mu/\nu$ with $\nu$ and $\mu \to \infty$. If $k = \mu/\nu$, from (6.2) it follows that the period of $\rho(\mathbf{r})$ along the $y$-axis is $D_2 = \nu d'_2 = 2\pi\nu/a'_2$. This consideration suggests an idea of representing the incommensurate phase as the limit of a sequence of long-period commensurate phases when $D_2 \to \infty$.

This idea, however, encounters a serious difficulty. In case $k = \mu/\nu$ while $\nu$ and $\mu$ change, one has in fact a devil's staircase (see [23] and references therein). Even if a variation of $k = \mu/\nu$ is arbitrarily small, symmetry and some physical quantities undergo irregular jumps [23], so that no definite limit exists as $\nu, \mu \to \infty$. For use later lets us point out the commensurate phases at the devil's staircase relevant to ammonium fluoberyllate seeing that these phases were not listed completely in [23]; besides, we choose the coordinate system as in Section 4. Inside the Brillouin zone, the irreducible representation $T_2$ analysed in Section 5 splits into two irreducible representations $T_2$ and $T_4$ under number $T31$ in Kovalev's tables [15]. It is unknown which of these two representations corresponds to curve 2 (the soft-mode branch) in Fig. 1(b). Because of this, we consider both the representations and list the point symmetry of the relevant commensurate phases (the symmetry of the third possible phase was not pointed out in [23]) implying that $K = ma_2/n$. For the representation $T_2$: (1) if $m/n$ = odd/even, one has point groups $C_{2v}$ with the $z$-polar axis, $C_{2h}$ and $C_s$; (2) if $m/n$ = odd/odd, one has point groups $C_{2v}$ with the $x$-polar axis, $C_{2h}$ and $C_s$; (3) if $m/n$ = even/odd, one has point groups $D_2$, $C_{2h}$ and $C_2$. For the representation $T_4$: (1) if $m/n$ = odd/even, one has point groups $C_{2v}$ with the $z$-polar axis, $C_{2h}$ and $C_s$; (2) if $m/n$ = odd/odd, one has point groups $D_2$, $C_{2h}$ and $C_2$; (3) if $m/n$ = even/odd, one has point groups $C_{2v}$ with the $x$-polar axis, $C_{2h}$ and $C_s$. This simple analysis that enables one to establish the point groups alone is sufficient for our purposes although a more sophisticated analysis analogous with the one carried out in Section 5 will give the relevant space groups as well. It may be remarked that the space groups listed in Table 1 on page 547 of [21] and in the table on page 85 of [22] correspond to the representation $T_4$ above (other coordinate systems are used in [21] and [22]). In order to avoid the symmetry jumps when changing $K$ and to have always the case where there is point group $C_{2v}$ with the $z$-polar axis as in the preceding section, one should so select the integers $m$ and $n$ that $m/n$ = odd/even irrespective of the representation, $T_2$ or $T_4$. In particular, when $m/n = 1/2$, one will have space group $C_{2v}^9$ of the preceding section.

Still another condition should be met. The above symmetries concern in fact the supercell with the period $D_2 = \nu d_2'$. When, according to Section 2, the incommensurate phase becomes domain-like (near the lock-in transition), the cells of period $d_2'$ inside the domains have the structure of the low-temperature commensurate phase, which amounts to saying that they are of $C_{2v}^9$ symmetry as well. The $C_{2v}^9$ symmetry transformations contain a translation by $\mathbf{d}_2'/2$, which signifies the translation by $\mathbf{D}_2/2 = \nu \mathbf{d}_2'/2$ for the supercell (the translations along the $x$- and $z$-axes are identical in the cell and supercell). If $\nu$ is odd ($\nu = 2\nu' + 1$), one has $\mathbf{D}_2/2 = \nu' \mathbf{d}_2' + \mathbf{d}_2'/2$, that is to say, one finds oneself after the translation in the middle of the cell as it should. Hence, the integer $\nu$ in $k = \mu/\nu$ must be odd. From Fig. 1 and the fact that $a_2' = a_2/2$ in (6.2) it follows that the fraction $m/n$ in $K = ma_2/n$ and $k = \mu/\nu$ are interrelated by $m/n = 1/2 - k/2 = (\nu - \mu)/(2\nu)$. Inasmuch as we should have $m/n$ = odd/even while $\nu$ is odd, the integer $\mu$ must be even. As a result, we see that, if the integer $\nu$ is odd while the integer $\mu$ is even in $k = \mu/\nu$, in the limit as $\mu$ and $\nu \to \infty$ we should arrive at the required incommensurate phase without any irregular jumps characteristic of the devil's staircase. In what follows we assume that the fraction $\mu/\nu$ is irreducible, which is equivalent to saying that the integers $\mu$ and $\nu$ are relatively prime.

We now turn to the long-period phases. Upon putting $k = \mu/\nu$ we apply the $C_{2v}^9$ symmetry operations to (6.2) and, analogously to (5.1), obtain

$$a_{lmnp} = a_{-l,-m,n,-p}(-1)^{l+\nu m+n} = a_{-l,m,n,p}(-1)^{\nu m} = a_{l,-m,n,-p}(-1)^{l+n}. \tag{6.4}$$

Here we have used the fact that $(-1)^{\mu p} = 1$ for $\mu$ is even. To this must be added the condition $a^*_{lmnp} = a_{-l-m-n-p}$ as $\rho(\mathbf{r})$ is real. Now, instead of the general form of (6.3), we are in position to write down the potential for the long-period phases which goes over into (5.8) if $k = 0$ (the fact that $\nu$ is odd is taken also into account):

$$U(\mathbf{r}) = \rho_0\sigma(0) + 4\cos a_1 x \sin a_3 z \sum_{p=0}^{\infty} \alpha_{1p}\sigma_{1p}\cos pka'_2 y + 4\cos a_3 z \sum_{p=-\infty}^{\infty} \alpha_{2p}\sigma_{2p}\sin(2+pk)a'_2 y$$

$$+ 4\sin a_1 x \sum_{p=-\infty}^{\infty} \alpha_{3p}\sigma_{3p}\sin(1+pk)a'_2 y, \qquad (6.5)$$

where $\alpha_{10} = ia_{1010}$, $\alpha_{1p} = 2ia_{101p}$ if $p \neq 0$, $\alpha_{2p} = ia_{021p}$, $\alpha_{3p} = -a_{110p}$, and

$$\sigma_{1p} = \sigma\left(\sqrt{a_1^2 + p^2k^2 a'^2_2 + a_3^2}\right), \quad \sigma_{2p} = \sigma\left(\sqrt{(2+pk)^2 a'^2_2 + a_3^2}\right), \quad \sigma_{3p} = \sigma\left(\sqrt{a_1^2 + (1+pk)^2 a'^2_2}\right). \quad (6.6)$$

Although the summation in (6.5) should not be extended up to infinity once the Kirkwood approximation is used, the potential of (6.5) contains a great many terms because the value of $k$ is usually small ($k \sim 0.01$ for ammonium fluoberyllate [17]).

As long as $k = \mu/\nu$, the potential $U(\mathbf{r})$ of (6.5) and thereby the density $\rho(\mathbf{r})$ in (3.1) are periodic and we can expand $\rho(\mathbf{r})$ in a Fourier series. We shall however obtain the coefficients $a_{lmn}$ of (4.1) whereas we need the coefficients $a_{lmnp}$ of (6.2). Therefore, an interrelation between $a_{lmn}$ and $a_{lmnp}$ is required. To this end we recast (4.1) for the long-period phases replacing $a_2$ by $a'_2/\nu$ since $D_2 = \nu d'_2$; besides, we substitute $s$ for $m$:

$$\rho(\mathbf{r}) = \sum_{l,s,n=-\infty}^{\infty} a_{lsn} e^{i(la_1 x + na_3 z) + isa'_2 y/\nu}. \qquad (6.7)$$

Comparing this with (6.2) where $k = \mu/\nu$ we see that $s = \nu m + \mu p$. This is a Diophantine equation for $m$ and $p$. As long as $\mu$ and $\nu$ are relatively prime, the equation admits a solution with arbitrary $s$; moreover, it has an infinite number of solutions with the same $s$. If one writes $s = \nu m' + \mu p'$ for another solution and compares this with the first one, one arrives at

$$\frac{\mu}{\nu} = \frac{m-m'}{p'-p}. \qquad (6.8)$$

The integers $\mu$ and $\nu$ being relatively prime, this relation is satisfied only if $m - m' = t\mu$ and $p' - p = t\nu$ with an arbitrary integer $t$, from which $m' = m - t\mu$ and $p' = p + t\nu$ for any other solution.

We see now that, to obtain Eq. (6.7) from (6.2), at a given $s$ we should add up all $a_{lmnp}$ for which $\nu m + \mu p = s$ with the result

$$a_{lsn} = \sum_{t=-\infty}^{\infty} a_{l,m-t\mu,n,p+t\nu}. \qquad (6.9)$$

The coefficients $a_{lmnp}$ in (6.1) should tend to zero as $|p|$ tends to infinity; the same occurs if $|m| \to \infty$ because $a_{lmn} \to 0$ in this limit. Consequently, as $\nu, \mu \to \infty$, only the term with $t = 0$ remains in the sum of (6.9), which finally yields

$$a_{lmnp} = a_{l,\nu m + \mu p, n}. \tag{6.10}$$

With this formula at our disposition we can calculate the Fourier coefficients $a_{lmn}$ with the help of the standard procedure, and we shall arrive at an equation of the type (3.8) into which the potential $U(\mathbf{r})$ of (6.5) should be substituted. Afterwards, with use made of Eq. (6.10) we shall find the following equations for the coefficients $\alpha_{ip}$ ($i = 1, 2, 3$) that figure in (6.5):

$$\frac{\alpha_{10}}{\rho_0} = \frac{J_{10}}{J_0}, \quad \frac{\alpha_{1p}}{\rho_0} = \frac{2J_{1p}}{J_0} \text{ if } p \neq 0, \quad \frac{\alpha_{2p}}{\rho_0} = \frac{J_{2p}}{J_0}, \quad \frac{\alpha_{3p}}{\rho_0} = \frac{J_{3p}}{J_0}, \tag{6.11}$$

where

$$J_0 = \frac{1}{8\pi^3} \int_0^{2\pi} \int_0^{2\pi} \int_0^{2\pi} e^{-U(\xi_1,\xi_2,\xi_3)} d\xi_1 d\xi_2 d\xi_3, \tag{6.12}$$

$$J_{1p} = \frac{1}{8\pi^3} \int_0^{2\pi} \int_0^{2\pi} \int_0^{2\pi} \cos\xi_1 \cos p\mu\xi_2 \sin\xi_3\, e^{-U(\xi_1,\xi_2,\xi_3)} d\xi_1 d\xi_2 d\xi_3, \tag{6.13}$$

$$J_{2p} = \frac{1}{8\pi^3} \int_0^{2\pi} \int_0^{2\pi} \int_0^{2\pi} \sin(2\nu + p\mu)\xi_2 \cos\xi_3\, e^{-U(\xi_1,\xi_2,\xi_3)} d\xi_1 d\xi_2 d\xi_3, \tag{6.14}$$

$$J_{3p} = \frac{1}{8\pi^3} \int_0^{2\pi} \int_0^{2\pi} \int_0^{2\pi} \sin\xi_1 \sin(\nu + p\mu)\xi_2\, e^{-U(\xi_1,\xi_2,\xi_3)} d\xi_1 d\xi_2 d\xi_3, \tag{6.15}$$

$$U(\xi_1, \xi_2, \xi_3) = 4\cos\xi_1 \sin\xi_3 \sum_{n=0}^{\infty} \beta_{1n} \cos n\mu\xi_2 + 4\cos\xi_3 \sum_{n=-\infty}^{\infty} \beta_{2n} \sin(2\nu + n\mu)\xi_2$$

$$+ 4\sin\xi_1 \sum_{n=-\infty}^{\infty} \beta_{3n} \sin(\nu + n\mu)\xi_2, \tag{6.16}$$

$$\beta_{1n} = \frac{\alpha_{1n}\sigma_{1n}}{\theta}, \quad \beta_{2n} = \frac{\alpha_{2n}\sigma_{2n}}{\theta}, \quad \beta_{3n} = \frac{\alpha_{3n}\sigma_{3n}}{\theta}. \tag{6.17}$$

From (6.12)–(6.15) it follows that

$$J_{1p} = -\frac{1}{4}\frac{\partial J_0}{\partial \beta_{1p}}, \quad J_{2p} = -\frac{1}{4}\frac{\partial J_0}{\partial \beta_{2p}}, \quad J_{3p} = -\frac{1}{4}\frac{\partial J_0}{\partial \beta_{3p}}. \tag{6.18}$$

The next step is to find out the form of the above integrals in the limit as $\nu$ and $\mu \to \infty$ in order to arrive at the incommensurate phase. In view of (6.18) it is sufficient to consider the integral $J_0$ alone, which is done in Appendix. As a result, we shall have for the incommensurate phase that

$$J_0 = \frac{1}{16\pi^4} \int_0^{2\pi} \int_0^{2\pi} \int_0^{2\pi} \int_0^{2\pi} e^{-U(\xi_1,\xi_2,\xi_3,\xi_4)} d\xi_1 d\xi_2 d\xi_3 d\xi_4, \tag{6.19}$$

$$U(\xi_1,\xi_2,\xi_3,\xi_4) = 4\cos\xi_1 \sin\xi_3 \sum_{n=0}^{\infty} \beta_{1n}\cos n\xi_4 + 4\cos\xi_3 \sum_{n=-\infty}^{\infty} \beta_{2n}\sin(2\xi_2 + n\xi_4)$$

$$+ 4\sin\xi_1 \sum_{n=-\infty}^{\infty} \beta_{3n}\sin(\xi_2 + n\xi_4). \tag{6.20}$$

It will be noted that $k$ has disappeared explicitly off (6.19) and (6.20); it remains only implicitly in $\beta_{in}$ through (6.6). The integrals $J_{1p}$, $J_{2p}$ and $J_{3p}$ can now be computed with the help of (6.18).

Combining (6.17) and (6.11) we arrive at a set of equations for $\beta_{ip}$:

$$\frac{\theta}{\rho_0 \sigma_1} = \frac{J_{10}}{\beta_{10} J_0}, \quad \frac{\theta}{\rho_0 \sigma_{1p}} = \frac{2 J_{1p}}{\beta_{1p} J_0} \text{ if } p \neq 0, \quad \frac{\theta}{\rho_0 \sigma_{2p}} = \frac{J_{2p}}{\beta_{2p} J_0}, \quad \frac{\theta}{\rho_0 \sigma_{3p}} = \frac{J_{3p}}{\beta_{3p} J_0}, \tag{6.21}$$

where $\sigma_1$ is the same as in (4.5). Having found the $\beta_{ip}$'s from the set as functions of the temperature $\theta$ we are able, by (6.11), to compute the coefficients $\alpha_{ip}$ that describe the incommensurate phase. It is necessary also to have the Helmholtz free energy. It can be calculated on a base of Eq. (4.27) of [4] upon substituting (6.2) and (6.3) there:

$$F = \theta N \ln\left[\frac{1}{\gamma}\left(\frac{m}{2\pi\theta}\right)^{3/2}\right] + \theta N \ln C - \frac{V}{2} \sum_{l,m,n,p} |a_{lmnp}|^2 \sigma\left(\sqrt{l^2 a_1^2 + (m+kp)^2 a_2'^2 + n^2 a_3^2}\right). \tag{6.22}$$

The constant $C$ herein, the same as in (3.1), is to be expressed in terms of the average density $\rho_0 = a_{000} = a_{0000}$ analogously to Eq. (3.10) of [4]. As a result, the first two terms on the right can be recast as in (3.10). For the example considered in the present section, one has

$$F = F_f - \theta N \ln J_0 - 2V\alpha_{10}^2 \sigma_1 - V\sum_{p=1}^{\infty}\alpha_{1p}^2\sigma_{1p} - 2V\sum_{p=-\infty}^{\infty}\alpha_{2p}^2\sigma_{2p} - 2V\sum_{p=-\infty}^{\infty}\alpha_{3p}^2\sigma_{3p}. \tag{6.23}$$

As distinct from the previous sections, the equations of (6.21) represent a great number of equations since the value of the integer $p$ can be very large in spite of the Kirkwood approximation employed, according to the remark concerning (6.5). It may be observed that the number of equations equal to two or three that we had in the previous sections (there was only one equation in [5]) is completely due to the use of the Kirkwood approximation. Without this approximation, we would have an infinite set of equations given by (3.8) even for a commensurate phase.

If $k = 0$, it is seen from (6.6) that $\sigma_{1p} = \sigma_{2p} = \sigma_1$ and $\sigma_{3p} = \sigma_3$ and thereby these quantities no longer depend on $p$. This being so, Eqs. (6.21) and (6.11) have a solution $\alpha_{ip} = 0$ if $p \neq 0$. Indeed, in the case of this solution the variable $\xi_4$ disappears from (6.20) and the integration over $\xi_4$ in (6.19) gives $2\pi$, so that $J_0 = I_0^{(2)}(\beta_1,\beta_2,\beta_3)$ of (5.11) upon putting $\beta_{i0} = \beta_i$. All integrals $J_{ip}$ with $p \neq 0$ vanish, which gives $\alpha_{ip} = 0$ if $p \neq 0$ in view of (6.11), whereas $J_{i0} = I_i^{(2)}(\beta_1,\beta_2,\beta_3)$ of (5.12–14). As a result, if $k = 0$, one will have the $C_{2v}^9$ phase treated in Section 5.

Considering the transition from the parent phase to the incommensurate one we set $\beta_{10} = \beta_{20} = \beta_1$, $\beta_{30} = 0$, and assume that only the first harmonics with the coefficients $\alpha_{31}$ and $\alpha_{3,-1}$ come into being initially because it is just these harmonics that correspond to the last term in (5.7). We expand the integrals $J_0$, $J_{31}$ and $J_{3,-1}$ in powers of $\alpha_{31}$ and $\alpha_{3,-1}$ and retain only linear terms. The result is placed in the last equation of (6.11) with account taken of (6.17). In this manner we obtain the following two equations

$$\left(\frac{\theta I_0^{(1)}(\beta_1,\beta_1)}{\rho_0} + 4\sigma_{31} I_{31}^{(2)}(\beta_1)\right)\alpha_{31} = 0, \quad \left(\frac{\theta I_0^{(1)}(\beta_1,\beta_1)}{\rho_0} + 4\sigma_{3,-1} I_{31}^{(2)}(\beta_1)\right)\alpha_{3,-1} = 0, \quad (6.24)$$

where $I_0^{(1)}(\beta_1,\beta_1)$ and $I_{31}^{(2)}(\beta_1)$ are the same integrals as in (4.8) and (5.21) and, according to (6.6),

$$\sigma_{31} = \sigma\left(\sqrt{a_1^2 + (1+k)^2 a_2'^2}\right), \quad \sigma_{3,-1} = \sigma\left(\sqrt{a_1^2 + (1-k)^2 a_2'^2}\right). \quad (6.25)$$

It follows from (6.24) that one of the coefficients $\alpha_{31}$ or $\alpha_{3,-1}$ acquires a nonzero value if the expression in the relevant brackets vanishes, which gives two temperatures

$$\frac{\theta}{\rho_0|\sigma_1|} = 4\frac{\sigma_{31} I_{31}^{(2)}(\beta_1)}{\sigma_1 I_0^{(1)}(\beta_1,\beta_1)}, \quad \frac{\theta}{\rho_0|\sigma_1|} = 4\frac{\sigma_{3,-1} I_{31}^{(2)}(\beta_1)}{\sigma_1 I_0^{(1)}(\beta_1,\beta_1)}. \quad (6.26)$$

If $k = 0$, then $\sigma_{31} = \sigma_{3,-1} = \sigma_3$, and both the temperatures coincide with the temperature of (5.22). The quantities $\sigma_{31}$ and $\sigma_{3,-1}$, however, cannot have an extremum at $k = 0$ because of the presence of terms linear in $k$. One of the quantities must be a maximum at $k \neq 0$ (for definiteness, we assume that $k > 0$). The maximum will yield the transition temperature to the incommensurate phase by virtue of (6.26), and the temperature will be higher than the one given by (5.22). Simultaneously, one will have the corresponding value of $k$. In the case of ammonium fluoberyllate the intermolecular potential is such that the value of $k$ will be small.

The above reasoning is akin to the one concerning Eq. (2.8). In the Landau theory, the behaviour of the soft-mode branch at $K = a_2/2$ displayed in Fig. 1(b) is reflected in the presence of the Lifshitz invariant (see Section 2). In the statistical approach where the Lifshitz invariant does not figure, that behaviour manifests itself in the presence of linear in $k$ terms in $\sigma_{ip}$ of (6.6).

Having found the phase transition temperature one can elucidate the further temperature behaviour of the incommensurate phase upon solving the equations of (6.21) (to do this, one must, of course, know the concrete form of $\sigma_{ip}$ that depends upon the intermolecular potential and the pair correlation function by (3.7) and (3.4)). The equations also contain $k$ via $\sigma_{ip}$. The equilibrium value of $k$ has to be computed by minimizing the Helmholtz free energy of (6.23) as far as it is implied in the present study that the external conditions are specified by the volume of the crystal (more precisely, by its periods) and by the temperature, so that the relevant thermodynamic potential is just the Helmholtz free energy.

One further point should be discussed. In the Landau theory, we had only two equations (2.5) and (2.6) that described the incommensurate phase (in the case of a two-component order parameter). In the present statistical theory, the incommensurate phase is described by a great many equations of (6.21) (although the equations are not differential). This is due to the following. In the Landau theory, one presumes that the order parameter is one of the normal coordinates while the normal coordinates are mutually independent in a linear approximation. In the statistical theory, the role of the order parameter is played by Fourier coefficients as mentioned at the outset of this section whereas different Fourier coefficients are interrelated. It may be added that the neglect of other normal coordinates below the phase transition point in the Landau theory is only an approximation whose validity is not clear (see the end of Section 2).

If the dispersion curves behave as the ones presented in Fig. 1(a), the quantities $\sigma_{ip}$ should depend upon $k$ quadratically as $\sigma_{1p}$ of (6.6). One can also deduce an equation equivalent to (6.26). In this case the quantity, say, $\sigma_{11}$ will have an extremum at $k = 0$. If, however, the extremum is not a maximum, the maximal possible value of $\sigma_{11}$ and the relevant temperature $\theta$ will anew be at $k \neq 0$, which amounts to saying that one will again have an incommensurate phase, of type II in this event. Hence, the present approach enables one to treat the type-II incommensurate phases as well.

## 7. Concluding remarks

In the present paper, a mathematical apparatus has been worked out which permits one to study commensurate as well as incommensurate phase transitions and the relevant phases from the viewpoint of statistical mechanics. It is shown how one can carry out the preliminary symmetry analysis when one deals with a concrete crystal. The analysis can be performed analogously with the one in the Landau phase transition theory where this analysis is well elaborated although some refinements are required which are not obligatory in the Landau theory. After the analysis one is able to deduce a set of equations that describe the emerging phases and corresponding phase transitions. The equations of the statistical approach contain quantities that may be directly calculated once the intermolecular potential and the pair correlation function are known while the equations of the Landau theory contain only phenomenological coefficients.

The Landau theory as applied to incommensurate phases leans heavily on the notion of the Lifshitz invariant. In the statistical approach, the Lifshitz invariant does not figure and the relevant behaviour of the soft-mode branch is reflected in another way. The treatment of an incommensurate phase in the statistical theory is substantially complicated because the symmetry of the phase cannot be described in terms of customary space groups. For this reason, a strategy of representing the incommensurate phase as the limit of a sequence of long-period commensurate phases whose period tends to infinity was adopted in the paper. The strategy, however, encounters a serious difficulty because a devil's staircase occurs in this situation. We have chosen a method of moving along the devil's staircase such that it was possible to avoid irregular jumps characteristic of the devil's staircase and to arrive at a definite limit. It should be added that the mathematical apparatus worked out in the paper can be used for study of incommensurate phases of type I as well as of type II.

The studies in the paper were carried out using displacive phase transitions as an example because a soft mode exists in this event and the occurring situation can be conveniently illustrated with the help of Fig. 1. All results of the paper remain valid for order-disorder phase transitions as well where occupation-modulated structures arise.

It should be emphasized that the present paper is a first, though essential, step in applying statistical mechanics to studying the incommensurate phases and the phase transitions relevant to them. We have shown that the phase transition temperature and properties of the emerging phase are expressible in terms of the function $\sigma(A)$ of (3.7) whose form depends upon the pair correlation function. Consequently, the next step is to find an equation for this last function leaning on the second equation of the BBGKY hierarchy. Seeing that the pair correlation function should be directly dependant on the intermolecular potential, the second step will enable one to express the phase transition temperature, peculiarities of the transition and properties of the incommensurate phase in terms of the intermolecular potential. At the same time to find the pair correlation function is not a simple matter inasmuch as the second BBGKY equation contains a triplet correlation function as well. The problem is complicated by the fact that, as distinct from a fluid, the pair correlation function in a crystal is anisotropic and its form is rather involved even in the case of an ordinary crystal [6].

Some remarks should be made as to the approach where one describes the symmetry of an incommensurate phase in terms of superspace groups [21, 22]. An especial investigation is needed in order to uniquely prescribe a superspace group for a given incommensurate phase, which requires elaborate techniques [22]. At the same time, in the Landau theory it suffices to know the space group of the parent phase and that of the low-temperature commensurate phase in order to write down unambiguously the free energy of the type (2.1) which enables one to investigate properties of the incommensurate phase. The same occurs in the present statistical approach where we have obtained a unique system of equations for the incommensurate phase in the crystal under study on a basis of the above space groups alone.

It is instructive nevertheless to compare the present approach and the superspace-group one. In line with the superspace-group approach we replace $y$ in the last term of the exponent of (6.2) by an independent variable, say, $u$ implying that we may always put $u = y$ at the end; for convenience we set $ka_2' = a_4$ as well. The same replacements are to be made in (6.3). Now (6.2) becomes an ordinary four-dimensional Fourier series for $\rho(\mathbf{r},u)$ whose coefficients $a_{lmnp}$ can be calculated in a standard way. With use made of (3.1) written for $\rho(\mathbf{r},u)$ and $U(\mathbf{r},u)$ we obtain a set of equations for $a_{lmnp}$ analogous to (3.8). To achieve a full analogy it needs to single out the term with $\sigma(0)$ and to determine the constant $C$ from the condition that $a_{0000} = \rho_0$. As a result, we arrive at the following set of equations

$$a_{lmnp} = \frac{\rho_0}{16\pi^4 G} \int_0^{2\pi}\int_0^{2\pi}\int_0^{2\pi}\int_0^{2\pi} e^{-\frac{1}{\theta}U_4(\xi_1,\xi_2,\xi_3,\xi_4)-i(l\xi_1+m\xi_2+n\xi_3+p\xi_4)} d\xi_1 d\xi_2 d\xi_3 d\xi_4, \quad (7.1)$$

where

$$G = \frac{1}{16\pi^4} \int_0^{2\pi}\int_0^{2\pi}\int_0^{2\pi}\int_0^{2\pi} e^{-\frac{1}{\theta}U_4(\xi_1,\xi_2,\xi_3,\xi_4)} d\xi_1 d\xi_2 d\xi_3 d\xi_4, \quad (7.2)$$

$$U_4(\xi_1, \xi_3, \xi_3, \xi_4) = \sum_{l,m,n,p=-\infty}^{\infty}{}' a_{lmnp}\, \sigma\!\left(\sqrt{l^2 a_1^2 + (m+kp)^2 a_1'^2 + n^2 a_3^2}\right) e^{i(l\xi_1+m\xi_2+n\xi_3+p\xi_4)}. \quad (7.3)$$

It may be noted that the auxiliary variable $u$ has disappeared off these formulae so that there is no need to put $u = y$ here.

If one specifies $U_4(\xi_1, \xi_3, \xi_3, \xi_4)$ of (7.3) on the basis of (6.5) with the above replacements, one will see that $G$ of (7.2) fully coincides with $J_0$ of (6.19). This result entails two important conclusions. First, the lengthy calculations of Appendix are accurate. Secondly, the strategy of long-period commensurate phases chosen in Section 6 and the method of overcoming the peculiarities of the devil's staircase have led to a correct limit.

At the same time, the set of equations (7.1) by itself is of little use. The example of the set of (3.8) analogous to (7.1) demonstrates that the set becomes meaningful and yields definite results only if one specifies the set for a concrete space group, which is seen from Sections 4 and 5 above and from Ref. [5]. However, even if one knows a definite superspace group for a given incommensurate phase and can specify the set of (7.1) with this superspace group, this will be of little avail for studying the relevant phase transitions. For example, the general form of the potential $U(\mathbf{r})$ of (5.2) for space group $C_{2v}^9$ says nothing about possible phase transitions. Only the potential $U(\mathbf{r})$ of (5.8) which is a particular case of (5.2) and which was deduced from the knowledge of the parent and low-temperature space groups enabled us to study the phase transitions. The potential $U(\mathbf{r})$ of (6.5) that was used for studying the incommensurate phase transition was again obtained with the help of those space groups alone, and no superspace groups were at all required.

**Appendix. Limit of the integral $J_0$ as $\nu$ and $\mu \to \infty$**

This limit concerns only the integration over $\xi_2$ and, instead of $J_0$ of (6.12), we shall consider the following integral with the same $U(\xi_1, \xi_2, \xi_3)$ as in (6.16):

$$I = \int_0^{2\pi} e^{-U(\xi_1,\xi_2,\xi_3)} d\xi_2 . \tag{A.1}$$

We change the variable of integration according to $\xi_2 = y/\nu$:

$$I = \frac{1}{\nu} \int_0^{2\pi\nu} e^{-U(y)} dy, \tag{A.2}$$

$$U(y) = 4\cos\xi_1 \sin\xi_3 \sum_{p=0}^{\infty} \beta_{1p} \cos pky + 4\cos\xi_3 \sum_{p=-\infty}^{\infty} \beta_{2p} \sin(2+pk)y$$

$$+ 4\sin\xi_1 \sum_{p=-\infty}^{\infty} \beta_{3p} \sin(1+pk)y, \tag{A.3}$$

where $k = \mu/\nu$ as before. We divide the interval of integration in (A.2) into $\nu$ intervals of length $2\pi$ and consequently represent the integral $I$ as a sum of $\nu$ integrals in each of which we change the variables of integration so that the integration shall be from 0 to $2\pi$. As a result,

$$I = \frac{1}{\nu} \sum_{j=0}^{\nu-1} \int_0^{2\pi} e^{-U_j(y)} dy, \tag{A.4}$$

$$U_j(y) = 4\cos\xi_1 \sin\xi_3 \sum_{p=0}^{\infty} \beta_{1p} \cos(2\pi jpk + pky) + 4\cos\xi_3 \sum_{p=-\infty}^{\infty} \beta_{2p} \sin[2\pi jpk + (2+pk)y]$$

$$+ 4\sin\xi_1 \sum_{p=-\infty}^{\infty} \beta_{3p} \sin[2\pi jpk + (1+pk)y]. \tag{A.5}$$

Instead of the discrete variable $j$, we introduce a continuous variable $t$ and consider the function (other variables are regarded as parameters)

$$f(t) = \exp\left\{-4\cos\xi_1\sin\xi_3 \sum_{p=0}^{\infty} \beta_{1p}\cos(2\pi pkt + pky) - 4\cos\xi_3 \sum_{p=-\infty}^{\infty} \beta_{2p}\sin[2\pi pkt + (2+pk)y]\right.$$

$$- 4\sin\xi_1 \sum_{p=-\infty}^{\infty} \beta_{3p} \sin[2\pi pkt + (1+pk)y] \Bigg\}. \tag{A.6}$$

This function is periodic with the period $d = 1/k$ and can be expanded into a Fourier series:

$$f(t) = \frac{a_0}{2} + \sum_{n=1}^{\infty} (a_n \cos 2\pi knt + b_n \sin 2\pi knt), \tag{A.7}$$

where

$$a_n = 2k \int_{-1/2k}^{1/2k} f(\tau)\cos 2\pi kn\tau \, d\tau = \frac{1}{\pi} \int_0^{2\pi} f\left(\frac{\xi}{2\pi k}\right) \cos n\xi \, d\xi, \tag{A.8}$$

the coefficient $b_n$ being of no importance for the present calculations.

The integral $I$ of (A.4) contains the sum

$$S = \sum_{j=0}^{\nu-1} f(j) = \frac{\nu a_0}{2} + \sum_{n=1}^{\infty} \sum_{j=0}^{\nu-1} (a_n \cos 2\pi knj + b_n \sin 2\pi knj). \tag{A.9}$$

Remembering that $k = \mu/\nu$, by virtue of Eqs. (1.342.1, 2) of [24] one has

$$\sum_{j=0}^{\nu-1} \cos 2\pi knj = \frac{\sin\pi\mu n}{\sin(\pi\mu n/\nu)} \cos[(\nu-1)\pi kn], \quad \sum_{j=0}^{\nu-1} \sin 2\pi knj = \frac{\sin\pi\mu n}{\sin(\pi\mu n/\nu)} \sin[(\nu-1)\pi kn]. \tag{A.10}$$

It is seen that these sums vanish owing to $\sin\pi\mu n$ except for the case where $n = p\nu$ with $p = 1, 2, 3, \ldots$ when one has 0/0. In this case, $\cos 2\pi p\mu j = 1$ while $\sin 2\pi p\mu j = 0$, and (A.9) yields finally

$$S = \frac{\nu a_0}{2} + \nu \sum_{p=1}^{\infty} a_{p\nu}. \tag{A.11}$$

The last sum should approach zero as $\nu \to \infty$ because the Fourier coefficients $a_n$ of (A.7) tend to zero as $n \to \infty$. In the event of a function of the type (A.6), the Fourier coefficients $a_n$ tend to zero exponentially inasmuch as the function has all derivatives and they are continuous.

Substituting all of these into (A.4) and letting $\nu \to \infty$, in the end we obtain

$$I = \frac{1}{2\pi} \int_0^{2\pi} dy \int_0^{2\pi} e^{-U(\xi)} d\xi, \tag{A.12}$$

$$U(\xi) = 4\cos\xi_1 \sin\xi_3 \sum_{p=0}^{\infty} \beta_{1p} \cos p(\xi + ky) + 4\cos\xi_3 \sum_{p=-\infty}^{\infty} \beta_{2p} \sin[2y + p(\xi + ky)]$$

$$+ 4\sin\xi_1 \sum_{p=-\infty}^{\infty} \beta_{3p} \sin[y + p(\xi + ky)]. \tag{A.13}$$

Exploiting the property embodied by Eq. (4.6) and placing the result in (6.12) we arrive at (6.19).